\documentclass[11pt,a4paper]{article}
\pdfoutput=1
\usepackage{a4wide}
\usepackage{jheppub}

\usepackage[T1]{fontenc} 

\usepackage{amsmath}
\usepackage{amssymb}
\usepackage{mathrsfs}
\usepackage{graphicx}
\usepackage[percent]{overpic}
\usepackage{latexsym}
\usepackage{bbm}
\usepackage{mathdots}

\newcommand{\be}{\begin{equation}}
\newcommand{\ee}{\end{equation}}
\renewcommand{\d}{\textrm{d}}

\newcommand{\SL}{\mathop{\rm SL}}
\newcommand{\GL}{\mathop{\rm GL}}
\newcommand{\SO}{\mathop{\rm SO}}
\newcommand{\CSO}{\mathop{\rm CSO}}
\newcommand{\SU}{\mathop{\rm SU}}
\newcommand{\e}{\mathrm{e}}
\newcommand{\tr}{\text{Tr}}
\newcommand{\dd}{\mathrm{d}}

\begin{document}

\title{Domain wall seeds in CSO-gauged supergravity}
\author{Juan Diaz Dorronsoro$^{a}$,}
\author{Harold Erbin$^b$,}
\author{Thomas Van Riet$^{a}$}
\affiliation{$^{a}$ Instituut voor Theoretische Fysica, K.U. Leuven,\\
	Celestijnenlaan 200D B-3001 Leuven, Belgium}
\affiliation{$^{b}$ CNRS, LPTENS, F-75231, Paris, France}

\emailAdd{juan.diaz@fys.kuleuven.be}
\emailAdd{erbin@lpt.ens.fr}
\emailAdd{thomas.vanriet@fys.kuleuven.be}

\abstract{Gravitational domain wall solutions in gauged supergravity are often constructed within truncations that do not include vectors. As a consequence the gauge group is only a global symmetry of this truncation. The consistency of the truncation requires the restriction to solutions with vanishing Noether charge under this global symmetry, since otherwise vector fields are sourced. We show that this has interesting consequences for the orbit structure of the solutions under the global symmetries. We investigate this for  CSO$(p,q,r)$-gaugings in various dimensions with scalar fields truncated to the SL$(n,\mathbb{R})/$SO$(n)$ subcoset. We prove that the seed solution -- which generates all other solutions using only global transformations -- has a diagonal coset matrix.  This means that there exists a transformation at the boundary of the geometry that diagonalises the coset matrix and that this same transformation also diagonalises the whole flow as a consequence of the vanishing charge.
}


\maketitle

\section{Introduction}
Supergravity theories can possess global symmetries of the equations of motion that manifest themselves as isometries of the scalar manifold. Typically there are more of those global symmetries if the number of supercharges is increased \cite{Freedman:2012zz}. For instance maximal (ungauged) supergravity in four dimensions enjoys an $E_{7(+7)}$ symmetry of the equations of motion, which corresponds to the isometry group of the 70-dimensional scalar manifold $E_{7(+7)}/\SU(8)$. These global symmetries are powerful in multiple ways.  They can be descendants of duality symmetries of the full string theory, allowing some insights into strongly coupled regimes of string theory \cite{Hull:1994ys}. More practical is their use as solution-generating symmetries. If the equations of motion are invariant under a group of symmetries $G$, then any solution to the equations can be mapped to another solution under the action of $G$. This leads to the concept of \emph{orbits} of solutions. 

Studying the orbits of solutions has been a very active field of research over the past decades in the context of black hole solutions to ungauged supergravity, see \cite{Breitenlohner:1987dg, Bertolini:1999je} for some original references. Less well studied is the orbit structure for domain wall solutions in gauged supergravity. 
This structure is expected to be somewhat more complicated since the symmetry group in gauged supergravity is not as big as in ungauged supergravity. The diminished symmetry is a consequence of the gauging which makes a subgroup of the isometry local and destroys the rest\footnote{Not necessarily all the rest of the global symmetry is destroyed. What is left is the normaliser of the gauge group in the original ungauged global symmetry group. We thank Adolfo Guarino for pointing this out.} of the original global symmetry by the introduction of a scalar potential (see for instance \cite{Trigiante:2016mnt}). 

Often one is interested in domain walls with maximally symmetric wall spaces and with all fields with non-trivial spin (apart from the metric) set to zero. The resulting equations of motion enjoy then the gauge group as a \emph{global} symmetry. For this to be consistent the scalar fields should not source any vector fields. This implies that the charge of the scalars has to vanish. This is a necessary and sufficient condition for the truncation to be valid and has to be checked separately when a solution of the scalar-metric system is found, since this truncation is not guaranteed to be consistent. We are therefore interested in understanding the orbit structure of solutions with zero charge under \emph{global} gauge rotations, which at first sight seems harder than the orbit structure under local rotations.  

In this paper we initiate a first systematic study of this in the context of a popular class of truncations of gauged supergravity theories in $D$ dimensions with $2<D<10$: the $\CSO(p,q,r)$-gaugings of theories with as scalar coset $\SL(n,\mathbb{R})/\SO(n)$ where $n=p+q+r$. A review of these theories, their solutions, and their link with (string theory) compactifications can be found in \cite{Roest:2004pk}. We prove here that all solutions in these theories can be found by transforming, under the \emph{global} symmetry group, the solutions with diagonal coset matrices. In other words, we show that there exists a rotation matrix inside the global symmetry group that diagonalises the coset matrix at the boundary and that this same matrix diagonalises the whole flow. This is non-trivial since a priori the transformation that diagonalises the matrix could depend on the position along the flow. It is also not evident that one can diagonalise the coset matrix using non-compact gauge groups. In this paper we prove both claims.

Since these statements can be confusing, let us consider a reasoning that can be found (indirectly stated) in the literature on susy domain walls for $\SO(n)$-gaugings, see for instance \cite{Cvetic:1999xx}. The argument is that prior to truncating the vectors, one can use the local rotation symmetry to diagonalise the coset matrix everywhere. Hence it is sufficient to look at solutions with diagonal matrices. These solutions do not source vectors since their $\SO(n)$-charge vanishes. The susy solution then obeys the typical flow equation
\begin{equation}\label{theflow}
\partial_z \phi^i \sim G^{ij}\partial_j W\,,
\end{equation}
where $\phi^i$ are the scalars, $G_{ij}$ the metric on field space, $W$ a function of the scalars called the real superpotential and $z$ is the coordinate transverse to the wall. For diagonal coset matrices only $n-1$ scalars are switched on and this equation is easily solved \cite{Cvetic:1999xx}. But imagine now that we started solving (\ref{theflow}) for a general coset matrix with all off-diagonal elements switched on and that the solutions have zero charge, as required for consistency. Then the resulting solution is expected to be diagonalisable using a local $\SO(n)$-transformation. Such a transformation will switch on vectors $A_{\mu}^I$ that are pure gauge, such that the field strengths still vanish. Hence one would expect the diagonal solution to solve the flow equation with $\partial_z$ replaced by a covariant derivative that includes the vectors. But since the charge vanishes the coupling to the diagonal scalars in the covariant derivative is absent. This can only be correct if it implies that the matrix that diagonalises the solution at a given value of $z$ diagonalises it for all $z$. This property is not at all obvious within the truncation and we explain in this paper how that works in detail. But our main result is that the same reasoning applies to the general $\CSO(p,q,r)$-gaugings. There is however an important difference between the $\SO(p,q)$-gaugings and their contracted counterparts $\CSO(p,q,r)$. In the latter case the global rotation that diagonalises the coset-matrix throughout the flow is bigger than the gauge group. So the rotation is not pure gauge.


\section{CSO-gauged SUGRA}

We consider a consistent truncation of gauged supergravity theories such that only the scalars $\phi^i$ ($i=1,\ldots N$) and the metric $g_{\mu\nu}$ are active. This subsystem is then described by the following action
\begin{equation}\label{actionOriginal}
S = \int \d^Dx ~\sqrt{|g|} \Bigl\{ \mathcal{R} -\tfrac{1}{2}G_{ij}(\phi)\partial\phi^i\partial\phi^j - V(\phi) \Bigr\}\,, 
\end{equation}
where $G_{ij}$ denotes the metric on the scalar manifold $\mathcal{X}$ and $V$ is the scalar potential. All maximal and half-maximal gauged supergravities allow a further consistent truncation of the scalars such that the scalar manifold $\mathcal{X}$ is
\begin{equation}\label{canonical}
\mathcal{X} = \frac{\SL(n,\mathbb{R})}{\SO(n)}\qquad \text{or}\qquad \mathcal{X} = \frac{\GL(n,\mathbb{R})}{\SO(n)}\,.
\end{equation}
For example in $\mathcal{N}=8, D=4$ maximal gauged SUGRA the scalar manifold is $E_{7(7)}/SU(8)$, describing 70 real scalars. Those 70 can be split into 35 scalars and 35 pseudo-scalars. The 35 scalars span $\SL(8,\mathbb{R})/\SO(8)$. Maximal supergravities in $D=7$ or higher do not need a truncation in order to be of the kind (\ref{canonical}) (or be products of such manifolds). 

We are particularly interested in $\CSO(p,q,r)$ gaugings as they form a ``canonical'' set of gaugings in maximal supergravity \cite{Roest:2004pk}. Well-known examples are the $\SO(n)$ gaugings originating from sphere compactifications of type II supergravity in $D=10$ or of $D=11$ supergravity. For instance $\SO(8)$ comes from $11D$ sugra on $S^7$, $\SO(6)$ from IIB on $S^5$, $\SO(5)$ from 11D sugra on $S^4$, $\SO(3)$ from IIA on $S^2$ or 11D on $S^3$ (as a group space, not as a coset). The $\SO(p,q)$ gauge groups can be found from generalised hyperbolic compactifications \cite{Hull:1988jw}.
 
Consider the coset representative $\mathcal{L}$ of $\SL(n,\mathbb{R})/\SO(n)$ in its fundamental representation. It is multiplied by the left with the $\SL(n,\mathbb{R})$ isometry transformations and from the right with the $\SO(n)$ isotropy group:
\be
\mathcal{L} \rightarrow \Omega \mathcal{L} h\qquad \Omega \in \SL(n,\mathbb{R})\,,\quad h \in \SO(n)\,.
\ee
It turns out useful to work with the symmetric coset element, $\mathcal{M} = \mathcal{L}\mathcal{L}^T$, which has unit determinant, is positive definite, and invariant under the isotropy group. Under the $\SL(n,\mathbb{R})$ isometry group it transforms as a bilinear form
\be
\mathcal{M}\quad \rightarrow\quad  \Omega \mathcal{M}\Omega^T\,.
\ee
The metric on $\SL(n,\mathbb{R})/\SO(n)$, which has $\SL(n,\mathbb{R})$, as an isometry group is given by
\be
G_{ij}\partial\phi^i\partial \phi^j = -\tfrac{1}{2}\text{Tr}[\partial \mathcal{M}\partial \mathcal{M}^{-1}]\,.
\ee
The coset can be extended to $\GL(n,\mathbb{R})/\SO(n)$ with an extra scalar $\varphi$ as follows
\be
G_{ij}\partial\phi^i\partial \phi^j = \partial\varphi\partial \varphi  -\tfrac{1}{2}\text{Tr}[\partial \mathcal{M}\partial \mathcal{M}^{-1}]\,.
\ee
The scalar potential for $\CSO$-gaugings can be written as
\be
V(\mathcal{M},\varphi) =  \e^{a\varphi} \Bigl[ \text{Tr}[\eta\mathcal{M}]^2 - \tfrac{1}{2} (\text{Tr}[\eta\mathcal{M}])^2 \Bigl]\,,
\ee
where $a$ is a number that depends on the dimension 
\be
a^2 = \frac{8}{n}-2\frac{D-3}{D-2}\,,
\ee
and the matrix $\eta$ is diagonal
\be \label{eta}
\eta = \begin{pmatrix}
\mathbbm{1}_{p\times p} & 0 & 0 \\
0 & -\mathbbm{1}_{q\times q} & 0 \\
0 & 0 & 0_{r\times r}
\end{pmatrix}\,,
\ee
where $p+q+r=n$. The combinations of $n$ and $D$ for which $a$ vanishes are also theories for which the scalar $\varphi$ is absent. 

The global symmetry group, in presence of the scalar potential, is the group that leaves the bilinear form $\eta$ invariant. When $r>0$ this is larger than the $\CSO(p,q,r)$-group, as we will explain below.

\section{Supersymmetric domain wall flows}
The metric ansatz for Minkowski-sliced domain wall solutions is
\be
\d s^2 =  f^2(z)\d z^2 + g^2(z)\d s^2_{D-1}\,,
\ee
where $\d s^2_{D-1}$ is the metric on Minkowski space. The warpfactors $g, f$ and the scalars $\phi^i$ are only functions of $z$. The warpfactor $f(z)$ is pure gauge and can be chosen freely. From the action \eqref{actionOriginal} one can derive the following effective action for domain-wall solutions:
\begin{align}\label{effectiveAction}
S_{eff}= \int \mathrm{d}z&\left[-\frac{1}{2}\dot{\phi}^2+\tfrac{1}{4}\tr(\dot{\mathcal{M}}^{-1}\dot{\mathcal{M}}) + (D-1)(D-2)\left( \frac{\dot{g}}{g} \right)^2  \right. \nonumber \\ 
&\phantom{m}\left.-  g^{2(D-1)}e^{a\phi} \tr(\eta \mathcal{M}\eta \mathcal{M})+\frac{g^{2(D-1)}}{2}e^{a\phi}[\tr(\eta \mathcal{M})]^2 \right]\,.
\end{align}
Varying \eqref{effectiveAction} gives the general second order equations of motion for $\mathcal{M}$:
\begin{equation}\label{2or-Equations}
\frac{1}{f g^{D-1}}\, \frac{\dd}{\dd z} \left(f^{-1} g^{D-1}\, \mathcal{M}^{-1} \dot{\mathcal{M}} \right)= \e^{a \varphi} \left(
			4 (\eta \mathcal{M})^2
			- 2 \tr(\eta \mathcal{M})\, \eta \mathcal{M}
			- \frac{4}{n} \tr(\eta \mathcal{M})^2
			+ \frac{2}{n}\, \big(\tr \eta \mathcal{M} \big)^2
			\right)~.
\end{equation}

Supersymmetric solutions solve first-order flow equations derived from a specific real superpotential $W$:
\begin{align}
& f^{-1} \dot{\phi}^i = -G^{ij}\partial_j W\,,\label{flow1}\\
& f^{-1} \frac{\dot{g}}{g} = \frac{1}{2(D-2)}W\,.
\end{align} 
The potential can be written in terms of $W$ as follows:
\be
V = \frac{1}{2}G^{ij}\partial_i W\partial_j W -\frac{(D-1)}{4(D-2)} W^2\,.
\ee
For the class of theories discussed here we have $
W  =  e^{a\varphi/2}\,\text{Tr}[\eta \mathcal{M}]$. 
Both $V$ and $W$ are manifestly invariant under the CSO gauge group. The flow equation for the scalars (\ref{flow1}) can be written as an equivalent first-order equation for the matrix $\mathcal{M}$ as follows:
\begin{align}
& f^{-1}e^{\frac{-a\varphi}{2}}\mathcal{M}^{-1}\dot{\mathcal{M}} =  -2\eta\mathcal{M} + \frac{2}{n}\text{Tr}{(\eta\mathcal{M})}\mathbbm{1}\,, \label{flow2}\\
& f^{-1}e^{\frac{-a\varphi}{2}} \dot{\varphi}=- \frac{a}{2}g^{D-1}\tr(\eta \mathcal{M})\,.
\end{align}
The matrix form of the first-order equations has not appeared earlier in the literature to our knowledge. The proof proceeds in the usual way by squaring the action. In the gauge $f=g^{D-1}$, we find that the following squared effective action,
\begin{align}
I_{eff}=&\int\d z\left\{(D-1)(D-2)\left(\frac{\dot{g}}{g}-\frac{g^{D-1}}{2(D-2)}e^{\frac{a\varphi}{2}}\tr(\eta \mathcal{M})\right)^2 -\frac{1}{2}\left[\dot{\varphi}+\frac{a}{2}g^{D-1}e^{\frac{a\varphi}{2}}\tr(\eta \mathcal{M})\right]^2 \right.\nonumber\\
&\phantom{mmmm}\left.-\frac{1}{4}\tr\left(\mathcal{M}^{-1}\dot{\mathcal{M}}+2g^{D-1}e^{\frac{a\varphi}{2}}\eta \mathcal{M}-\frac{2g^{D-1}}{n}e^{\frac{a\varphi}{2}}\tr(\eta \mathcal{M})\mathbbm{1}\right)^2\right\}~, \label{Ieff}
\end{align}
is equivalent to \eqref{effectiveAction}.

The diagonal solution was constructed in \cite{Bergshoeff:2004nq} (see also \cite{Cvetic:1999xx, Cvetic:2000zu}) in the gauge $f=g^{3-D}$ and is given by\footnote{In the case $n=3$ the solution with all axions turned on was found in \cite{AlonsoAlberca:2003jq}, but we do not require it here since we only care about seed solutions.}:
\begin{align}
&\d s^2 = h^{(3-D)/(2D-4)}\d z^2 + h^{1/(2D-4)}\d s^2_{D-1}~, \label{diag1}\\
& e^\varphi =h^{-a/4}~, \label{diag2}\\
& \mathcal{M}= h^{1/n}\text{diag}(1/h_1,\ldots,1/h_n)\,,\label{diag3}
\end{align}
where
\be
h_i = 2\eta_{i}z + \ell_i^2\,,\qquad h=h_1\ldots h_n\,, 
\ee
with $\eta_i$  the $i$'th diagonal entry of $\eta$ (hence $\pm 1$ or $0$) and $\ell_i$ an arbitrary integration constant. In the context of maximal supergravity, these flows were shown to preserve half of the supersymmetries \cite{Bergshoeff:2004nq,Cvetic:1999xx, Cvetic:2000zu}.

In the next sections, we will argue that all susy Minkowski-sliced domain wall solutions are related to this diagonal solution through a global transformation. The same also holds for all non-susy solutions with vanishing Noether charges: the seed will again be diagonal.

\section{Noether charges}

The effective action \eqref{effectiveAction} for finding domain walls 
corresponds to a normal Hamiltonian system in classical mechanics with the spatial coordinate $z$ playing the role of time. Since the effective action has the gauge group $G=CSO(p,q,r)$ as a global symmetry, there must be a set of conserved charges equal to the dimension of $G$, which remain constant along the flow in $z$. These conserved charges can always be written in terms of momenta and generalised coordinates and hence provide a set of first-order equations. We now show that for  supersymmetric solutions in our models these first-order equations contain no new information since they are implied by (\ref{flow2}). 

To gain some intuition we first consider the case without potential. The equations of motion for the scalars in the gauge $f=g^{D-1}$ decouple from the metric and are derived from the geodesic action
\be
S_g = \int \d z\, G_{ij}\dot{\phi}^i\dot{\phi}^j\,.
\ee
The geodesic equations for the scalars in $\mathcal{M}$ are summarized as
\be
\frac{\d}{\d z}(\mathcal{M}^{-1}\dot{\mathcal{M}}) = 0 \,.
\ee
This can be integrated once 
\begin{equation}\label{geodesicNoether}
\mathcal{M}^{-1}\dot{\mathcal{M}} = Q\,,
\end{equation}
with $Q$ any traceless matrix. This equation is further integrated as
\be
\mathcal{M}(z) = \mathcal{M}(0)e^{Qz}\,.
\ee
So remarkably all geodesics on $\SL(n,\mathbb{R})/\SO(n)$ can be found in the language of the symmetric coset matrix in a trivial manner\footnote{This is in sharp contrast with the algorithms that are developed to integrate the equations at the level of the scalars \cite{Fre:2003ep, Chemissany:2009hq, Chemissany:2009af} where the solutions can be found algorithmically but the computations and expressions are complicated.}. From (\ref{geodesicNoether}) one finds that $Q$ is in the Lie algebra of $\SL(n,\mathbb{R})$ since it is traceless and corresponds to the matrix of Noether charges.

Once we deform the geodesic motion by the potential $V$, only those Noether charges corresponding to the symmetry preserved by $V$ should survive. Hence the Noether charge matrix corresponding to the CSO gauged supergravity theories  should be related to the projection of the general $Q \in Lie(\text{SL}(n))$ to $Lie(\text{CSO}(p,q,r))$. An explicit computation of the Noether charge carried out in appendix \ref{App:Noether} yields
\be\label{Noether}
\mathcal{Q}=\eta\dot{\mathcal{M}}\mathcal{M}^{-1}-\mathcal{M}^{-1}\dot{\mathcal{M}}\eta~.
\ee 
Substituting the first-order equation (\ref{flow2}) into (\ref{Noether}) one finds that supersymmetric solutions have vanishing charge.  Hence the supersymmetric solutions source no vectors, consistent with our truncation. Moreover, one can explicitly check that $\mathcal{Q}\eta$ (which is also conserved) is an element of the CSO algebra.

We already explained in the Introduction that we expect solutions with zero Noether charges to behave in a special way under global transformations of the gauge group. Namely a solution whose coset matrix can be diagonalised at some value for $z$ using a gauge transformation will be diagonal for all values of $z$.  
We have included a computationally explicit proof of this in Appendix \ref{App:orthogonal} for $\SO(p,q)$-gaugings. For gaugings with contracted gauge groups the story complicates somewhat in the sense that a bigger transformation is needed to diagonalise a solution. But again the vanishing of the Noether charge will turn out sufficient to argue that all solutions are global transformations of the solution with diagonal coset matrix.

The reasoning uses the equations of motion \eqref{2or-Equations}, which can be schematically written as
\begin{equation}\label{HJflow}
\ddot{\mathcal{M}} =  F(\mathcal{M})\,.
\end{equation} 
Since the matrix $F(\mathcal{M})$ is diagonal for diagonal $\mathcal{M}$, if both $\mathcal{M}$ and its first derivative are diagonal at some point, the whole flow will remain diagonal. We will show that if we require the Noether charge to vanish, we can diagonalize $\mathcal{M}$ and $\dot{\mathcal{M}}$ at a reference value (e.g.~$z=0$) using a transformation inside the global symmetry group of the action. This ensures that all solutions with vanishing Noether charge (and hence all the solutions consistent with the truncation) can be obtained through a global symmetry transformation from the diagonal seed solutions.

The intuition behind this is simple. Consider for instance a spherically symmetric system in classical mechanics for which the angular momenta vanish. Since angular momentum is linked to rotation, there exists a frame in which the system has no rotation and all angular variables vanish at all times. Let us now extend this logic to our context.  If it is possible to diagonalise $\mathcal{M}(0)$ using $\SO(p,q)$ then the general coset element can be written as $\mathcal{L} = \mathcal{P} \times \mathcal{D}$, with $\mathcal{D}$ diagonal and $\mathcal{P}$ in $\SO(p,q)$. The degrees of freedom are explicitly separated into `radii' sitting in $\mathcal{D}$ and `angular variables' residing in $\mathcal{P}$. From the point of view of the scalar potential the variables $\mathcal{P}$ are cyclic.  If the variables in $\mathcal{P}$ are then zero in a certain frame (basis) at $z=0$ they will remain zero throughout the flow. Hence the transformation that brings one to the frame of vanishing angles at the boundary brings us to vanishing angles along the flow.

\section{Normal forms} \label{sec:normal2}

We now set out to prove the main claim made in the introduction that all solutions of the $\CSO$-gaugings with zero charge can be found performing a global transformation on solutions with diagonal coset matrices.  We first discuss $\SO(p,q)$ gaugings and then we treat the contracted $\CSO$ algebras.

\subsection{Normal form for \texorpdfstring{$\SO(p,q)$}{} gaugings}
We  start by arguing that any coset matrix $\mathcal{M}$ evaluated at the reference value $z=0$ can be diagonalised using an $\SO(p,q)$ transformation. Using the result form the previous section this then implies that the solution will be diagonal globally.

It is well-know that $\mathcal{M}(0)$ can be diagonalised using an $\SO(n)$-transformation since it is a symmetric bilinear form. We want to generalise this to $\SO(p,q)$-transformations with $p+q=n$. The statement is not true for general symmetric bilinear forms, but relies on $\mathcal{M}(0)$ being positive definite. Our proof uses a detour via the normal form of the coset element of $\SL(n,\mathbb{R})/\SO(p,q)$ constructed in \cite{Normal, Bergshoeff:2008be}. To make this connection, note that any symmetric matrix $\mathcal{M}$ can be written as 
\be\label{Adef}
\mathcal{M} = \mathcal{A} \eta\,,
\ee
with $\eta$ the invariant bilinear form under $\SO(p,q)$ defined earlier and $\mathcal{A}$ some \emph{generalised symmetric} matrix, which means
\be
\mathcal{A}^T = \eta \mathcal{A} \eta\,.
\ee
Such a matrix $\mathcal{A}$ can be regarded as being in the Lie algebra of $\SL(n,\mathbb{R})/\SO(p,q)$. It was shown in \cite{Normal, Bergshoeff:2008be} that one can bring $\mathcal{A}$ into its Jordan normal form $\mathcal{J}$ through a similarity transformation $\Omega\in\SO(p,q)$, so that
\begin{equation}
\mathcal{A}=\Omega \mathcal{J} \Omega^{-1}~.
\end{equation}
The proof of this proceeds by showing that one can always find a basis for $\eta$ in which the normal form of a generalised symmetric matrix is itself generalised symmetric: $\mathcal{J}^T=\eta \mathcal{J}\eta~$. This means that the similarity transformation is inside $\SO(p,q)$. The detailed proof of this statement can be found in \cite{Normal, Bergshoeff:2008be}. In Appendix \ref{App:Jordan} we show that, due to $\mathcal{M}$ being positive definite, the normal form $\mathcal{J}$ has to be strictly diagonal. After this we get
\be
\mathcal{M} = \mathcal{A} \eta=\Omega \mathcal{J} \Omega^{-1}\eta=\Omega \mathcal{J}\eta \Omega^{T}\,,
\ee
and we therefore see that we can indeed diagonalize $\mathcal{M}$ with an $\SO(p,q)$ transformation.

\subsection{Normal form for \texorpdfstring{$\CSO(p,q,r)$}{} gaugings}\label{normalCSO}
For the $\CSO(p,q,r)$ gaugings the global symmetry group is larger than the gauge group. The $\CSO$ group can be obtained by exponentiating the Lie algebra elements $g_{ij}$ given by \cite{Roest:2004pk}
\begin{equation}
(g_{ij})^k_{\phantom{k}l}=\delta_{[i}^k\eta_{j]l}~,
\end{equation}
whereas the global symmetries of the action are larger and contain all the transformations $\Gamma$ that leave $\eta$ invariant
\be\label{etaInvariant}
\eta= \Gamma^T\eta \Gamma\,.
\ee
This second group is larger, since even though one has
\begin{equation}
\exp\left(\lambda^{ij}g_{ij}^T\right)\eta\exp\left(\lambda^{ij}g_{ij}\right)=\eta
\end{equation}
for any choice of constants $\lambda^{ij}$, not all the matrices $\Gamma$ which leave $\eta$ invariant are of the form $\exp\left(\lambda^{ij}g_{ij}\right)$. Indeed, \eqref{etaInvariant} only implies that
\begin{equation}
\Gamma=\begin{pmatrix}
\Omega & 0_{(p+q)\times r} \\ 
\mathcal{A}_{r\times(p+q)} & \mathcal{B}_{r\times r}
\end{pmatrix}~,
\end{equation}
where $\Omega\in\text{SO}(p,q)$, $\mathcal{A}$ is an arbitrary $r$ by $(p+q)$ matrix and $\mathcal{B}$ is an arbitrary $r$ by $r$ matrix.

We will argue now that we can diagonalize any susy flow through a global $\Gamma$ transformation. To this aim, let us split $\mathcal{M}(0)$ into blocks as follows:
\begin{equation}
\mathcal{M}(0)=\begin{pmatrix}
\mathcal{M}_1(0) & \mathcal{M}_2(0) \\ 
\mathcal{M}_2^T(0) & \mathcal{M}_3(0)
\end{pmatrix} ~.
\end{equation}
Here, $\mathcal{M}_1$ has size $(p+q)\times(p+q)$ and $\mathcal{M}_3$ has size $r\times r$. From the arguments of the previous section, we know that there is a matrix $\Omega\in\text{SO}(p,q)$ such that $\Omega\mathcal{M}_1(0)\Omega^T=\mathcal{D}_1$ is diagonal. Let us transform $\mathcal{M}$ with the matrix
\begin{equation}
\Gamma=\begin{pmatrix}
\Omega & 0_{(p+q)\times r} \\ 
-\mathcal{B}\mathcal{M}_2^T(0)\mathcal{M}_1(0)^{-1} & \mathcal{B}
\end{pmatrix}~,
\end{equation}
where for the moment we keep $\mathcal{B}$ arbitrary. Then we get
\begin{equation}
\Gamma\mathcal{M}(0)\Gamma^T=\begin{pmatrix}
\mathcal{D}_1 & 0_{(p+q)\times r} \\ 
0_{r\times(p+q)} & \mathcal{B}\mathcal{S}\mathcal{B}^T
\end{pmatrix}~,\quad \quad \mathcal{S}=\mathcal{M}_3(0)-\mathcal{M}_2^T(0)\mathcal{M}_1^{-1}(0)\mathcal{M}_2(0)~.
\end{equation}
The matrix $\mathcal{S}$ is symmetric, and moreover it is positive definite.\footnote{The way to see this last point is by remarking that $\mathcal{S}$ is the combination that appears on the lower-right block of the inverse matrix $\mathcal{M}(0)^{-1}$. Indeed,
\begin{equation}
\mathcal{M}(0)^{-1}=\begin{pmatrix}
(\cdots) & (\cdots) \\ 
(\cdots) & \mathcal{S}^{-1}
\end{pmatrix}
\end{equation}
whenever $\mathcal{S}$ is not singular. The matrix $\mathcal{S}$ is also known as the Schur complement of $\mathcal{M}_1(0)$, and the matrix $\mathcal{M}(0)$ is positive definite if and only if $\mathcal{M}_1(0)$ and its Schur complement are both positive definite.}
From the properties of $\mathcal{S}$ we know that there exists an $\SO(r)$ transformation $\mathcal{O}$ which brings it into a diagonal form: 
\begin{equation}
\mathcal{O}\mathcal{S}\mathcal{O}^T=\text{diag}(d_1,\ldots,d_r)~,
\end{equation}
with $d_i>0$. If we then let
\begin{equation}
\mathcal{B}=\text{diag}\left(\frac{1}{\sqrt{d_1}}~,\ldots,\frac{1}{\sqrt{d_r}}\right)\cdot\mathcal{O}~,
\end{equation}
we see that
\begin{equation}
\mathcal{B}\mathcal{S}\mathcal{B}^T=\mathbbm{1}_r~.
\end{equation}
This shows that using the $\Gamma$ transformation we just described, we get
\begin{equation}
\Gamma\mathcal{M}(0)\Gamma^T=\begin{pmatrix}
\mathcal{D}_1 & 0_{(p+q)\times r} \\ 
0_{r\times(p+q)} & \mathbbm{1}_r
\end{pmatrix}~.
\end{equation}

We now argue why this transformation will diagonalise the coset matrix at all values for $z$ if the Noether charge vanishes. This result is not implied from the results in the previous section since we only require that the Noether charge under $\CSO$ rotations are zero, whereas the $\Gamma$-transformation above is typically inside a larger symmetry group. Hence we have to prove that the $\Gamma$-transformation will diagonalise the solution for all values of $z$.

Suppose therefore that $\mathcal{M}$ is a solution of the second order equations of motion such that
\begin{equation}
\mathcal{Q}=\eta\dot{\mathcal{M}}\mathcal{M}^{-1}-\mathcal{M}^{-1}\dot{\mathcal{M}}\eta=0~.
\end{equation}
Let us expand such a solution around the reference value $z=0$ as
\begin{equation}
\mathcal{M}=\mathcal{M}(0)+\dot{\mathcal{M}}(0)~z+\mathcal{O}(z^2)~.
\end{equation}
After performing a $\Gamma$ transformation as the one we described, we can write
\begin{equation}
\mathcal{M}=\begin{pmatrix}
\mathcal{D}_1 & 0 \\ 
0 & \mathbbm{1}_r
\end{pmatrix}+\begin{pmatrix}
\dot{\mathcal{M}}_1(0) & \dot{\mathcal{M}}_2(0) \\ 
\dot{\mathcal{M}}_2^T(0) & \dot{\mathcal{M}}_3(0)
\end{pmatrix}~z+\mathcal{O}(z^2)~.
\end{equation}
The Noether charge is therefore
\begin{equation}
\mathcal{Q}=\begin{pmatrix}
\eta_s\dot{\mathcal{M}}_1(0)\mathcal{D}_1-\mathcal{D}_1\dot{\mathcal{M}}_1(0)\eta_s & \eta_s\dot{\mathcal{M}}_2(0) \\ 
-\dot{\mathcal{M}}_2^T(0)\eta_s & 0
\end{pmatrix}~,
\end{equation}
where $\eta_s$ is a diagonal matrix with $p$ times $+1$ and $q$ times $-1$. If this charge vanishes, we know that $\dot{\mathcal{M}}_2(0)$ must vanish, and therefore we have
\begin{equation}
\mathcal{M}=\begin{pmatrix}
\mathcal{D}_1 & 0 \\ 
0 & \mathbbm{1}_r
\end{pmatrix}+\begin{pmatrix}
\dot{\mathcal{M}}_1(0) & 0 \\ 
0 & \dot{\mathcal{M}}_3(0)
\end{pmatrix}~z+\mathcal{O}(z^2)~.
\end{equation}
In this frame, the equations of motion for the upper-left diagonal part of the matrix $\mathcal{M}$ are precisely the same equations of motion of an $\SO(p,q)$ gauging, and we can use the same arguments of the previous section to show that this part of $\mathcal{M}$ will remain diagonal throughout the flow. This implies that $\dot{\mathcal{M}}_1(0)=\dot{\mathcal{D}}_1$ is of course diagonal.

As for the lower-right component of $\mathcal{M}$, we remark that since $\dot{\mathcal{M}}_3(0)$ is symmetric, we can diagonalize it through an $\SO(r)$ transformation $\mathcal{O}_r$ such that $\mathcal{O}_r\dot{\mathcal{M}}_3(0)\mathcal{O}_r^T=\dot{\mathcal{D}}_3$. Transforming now $\mathcal{M}$ as 
\begin{equation}
\mathcal{M}\rightarrow \begin{pmatrix}
\mathbbm{1}_{p+q} & 0 \\ 
0 & \mathcal{O}_r
\end{pmatrix} \mathcal{M}
\begin{pmatrix}
\mathbbm{1}_{p+q} & 0 \\ 
0 & \mathcal{O}_r^T
\end{pmatrix}~,
\end{equation}
we get around $z=0$
\begin{equation}
\mathcal{M}\rightarrow\begin{pmatrix}
\mathcal{D}_1 & 0 \\ 
0 & \mathbbm{1}_r
\end{pmatrix}+\begin{pmatrix}
\dot{\mathcal{D}}_1 & 0 \\ 
0 & \dot{\mathcal{D}}_3
\end{pmatrix}~z+\mathcal{O}(z^2)~.
\end{equation}
After applying the global transformations we described above, we have shown that both the coset matrix and its first derivative can be made diagonal at the reference value $z=0$. Due to the structure of the second order equations of motion, the matrix will preserve its diagonal form along the whole flow. This concludes the proof that any consistent solution (therefore with vanishing Noether charge) can be brought into a diagonal form through a global symmetry transformation.


\section{Discussion} \label{concl}

We have investigated the space of domain wall flows in $\CSO(p,q,r)$-gauged supergravities with scalar fields truncated to the coset $\SL(n,\mathbb{R})/\SO(n)$ (or $\GL(n,\mathbb{R})/\SO(n)$), where $n=p+q+r$. We have emphasized how the gauge group is a global and not a local symmetry once the vectors are truncated and that this requires the $\CSO(p,q,r)$-charge to be zero. We then showed that all the zero-charge solutions of the theory can be found by letting the global symmetry act on solutions with diagonal coset matrices. As the vanishing of the Noether charge is a necessary consistency condition for any solution of the truncated theory, we have proved that all the consistent solutions of CSO gaugings (within our truncation) can be obtained through a global rotation of a diagonal solution. All supersymmetric domain walls with diagonal coset matrices were constructed in  \cite{Cvetic:1999xx, Cvetic:2000zu, Bergshoeff:2004nq, Roest:2004pk}. For supersymmetric flows the results we found were implicitly known for $\SO(n)$ gaugings. For all other $\CSO(p,q,r)$-gaugings our findings were to our knowledge not explained in the literature\footnote{Aside some comments in \cite{AlonsoAlberca:2003jq} about the case $n=3$ where the claim can be found that ``the change of $\SL(n,\mathbb{R})$-frame'' diagonalises the solution.}. We emphasize that this result is more surprising for the gaugings with contracted gauge groups ($r>0$). In that case the global transformation that diagonalises the flows are outside of the gauge group, but still within the global symmetry of the action. In the extreme case $r=n$, there is no gauging and hence no scalar potential so that the flows describe geodesic curves on $\SL(n,\mathbb{R})/\SO(n)$. Those curves were known to be diagonalisable using $\SL(n,\mathbb{R})$ (see for instance \cite{Bergshoeff:2008be}). The reason the flow can be diagonalised when $0<r<n$ is a mixture between the different reasonings used for the extremes $r=0$ and $r=n$.

This work is a first step towards classifying (supersymmetric) domain wall flows in gauged supergravity. To achieve that goal, one should go beyond the $\SL(n,\mathbb{R})/\SO(n)$-truncation\footnote{Another extension could be to look at the $\omega$-deformations of the standard gaugings \cite{DallAgata:2012mfj} within the $\SL(n,\mathbb{R})/\SO(n)$-truncation.  Using the results of \cite{Guarino:2015tja} it seems that the findings of this paper still go through such that again the diagonal solutions are the seed solutions.}. We then expect that  a very non-trivial orbit structure should arise. 

\section*{Acknowledgements}
We thank Nikolay Bobev, Fridrik Gautason  for useful discussions and especially Adolfo Guarino and Mario Trigiante for discussions and essential feedback on a earlier draft. The work of JDD and TVR is supported by the FWO odysseus grant G.0.E52.14N and by the C16/16/005 grant of the KULeuven. We furthermore acknowledge support from the European Science Foundation Holograv Network  and the COST Action MP1210 `The String Theory Universe'.

\appendix

\section{Noether currents} \label{App:Noether}

Consider the effective action \eqref{effectiveAction}.
It is invariant under global $\CSO(p,q,r)$ transformations $\mathcal{M}\rightarrow\Omega \mathcal{M}\Omega^T$, with $\Omega$ an element of the $\CSO(p,q,r)$ group such that $\Omega^T\eta\Omega=\eta$. Infinitesimally we can write
\begin{equation}\label{NoetherGenerator}
\Omega=\mathbbm{1}+\lambda^{ij}g_{ij}+\mathcal{O}(\lambda^2)~.
\end{equation}
Here, $g_{ij}$ label the generators of the algebra and $\lambda^{ij}$ are transformation parameters, which are taken to be infinitesimal. In order to find the conserved charges we allow $\lambda^{ij}=\lambda^{ij}(z)$ and follow the standard Noether procedure. To simplify the notation, we write $h\equiv\lambda^{ij}g_{ij}$. The two last terms of the action remain invariant even when $h=h(z)$, so that we only consider the kinetic term and find:
\begin{align}
S_{eff}\rightarrow & S_{eff} - \frac{1}{2}\int\d z~\tr\left(\dot{\mathcal{M}} \mathcal{M}^{-1}\dot{h}+\mathcal{M}^{-1}\dot{\mathcal{M}}\dot{h}^T\right)+\mathcal{O}(\lambda^2)\\
=& S_{eff} - \int \d z~\tr\left(\dot{\mathcal{M}} \mathcal{M}^{-1}\dot{h}\right)+\mathcal{O}(\lambda^2)~.
\end{align}
Integrating by parts, throwing away a total derivative and by the standard argument of letting $\lambda^{ij}$ be constant, we get a set of conserved charges
\begin{equation}
\mathcal{Q}_{ij}=\tr\left(\dot{\mathcal{M}} \mathcal{M}^{-1}g_{ij}\right)~,\quad \dot{\mathcal{Q}}_{ij}=0~.
\end{equation}
The generators of the CSO group are \cite{Roest:2004pk}
\begin{equation}
(g_{ij})^k_{\phantom{k}l}=\delta_{[i}^k\eta_{j]l}~.
\end{equation}
A short computation then reveals that
\begin{equation}
\mathcal{Q}=\eta\dot{\mathcal{M}}\mathcal{M}^{-1}-\mathcal{M}^{-1}\dot{\mathcal{M}}\eta~.
\end{equation}

\section{Normal forms and vanishing Noether charges} \label{App:orthogonal}

Here we prove explicitly for SO($p,q$) gaugings that the matrix that diagonalises $\mathcal{M}$ at $z=0$ diagonalises $\mathcal{M}$ throughout the flow if the Noether charge $\mathcal{Q}$ vanishes. Inserting
\begin{equation}
	\mathcal{M}= \Omega \mathcal{D}\eta\Omega^T
\end{equation} 
into \eqref{Noether} and requiring $\mathcal{Q}=0$ yields
\begin{equation}
	\label{dw:eq:1st-order-A}
	2 \omega = \mathcal{D}\, \omega \mathcal{D}^{-1} + \mathcal{D}^{-1}\, \omega \mathcal{D}, \qquad
	\omega = \Omega^{-1} \dd\Omega \in \text{so}(p, q).
\end{equation}
If we let $\mathcal{D}=\text{diag}(d_1,\ldots,d_n)$, the previous equations becomes in components
\begin{equation}\label{componentsOmega}
2\omega_{ij}=\omega_{ij}\left(\frac{d_i}{d_j}+\frac{d_j}{d_i}\right)~.
\end{equation}
We recall that for $\omega\in\text{so}(p, q)$, all the diagonal elements of $\omega$ vanish. If all the eigenvalues of $\mathcal{M}$ are different (so that $d_i\neq d_j$ for all $i,j$), then \eqref{componentsOmega} further implies that $\omega_{ij}=0$, so that $\omega=0$ and
\begin{equation}
	\dd\Omega = 0
	\quad \Longrightarrow \quad
	\Omega = \mathrm{cst}.
\end{equation} 

What happens when some of the eigenvalues of $\mathcal{M}$ are equal? Suppose for instance that we are dealing with the extreme case in which all the eigenvalues are the same, so that also $d_i=d_j$ for all $i,j$. Then we get
\begin{equation}
\mathcal{M}=\Omega \mathcal{D}\eta\Omega^T=\Omega \mathcal{D}(\Omega^T)^{-1}\eta=\mathcal{D}\eta~,
\end{equation}
so that $\mathcal{M}$ is diagonal for any $\Omega$, and there is no need for diagonalizing it in the first place. We could then trivially keep $\Omega$ constant in this case too.

The argument for the constancy of $\Omega$ when only some of the eigenvalues of $\mathcal{M}$ are equal, follows from realizing that the only components $\omega_{ij}$ which can differ from zero are those for which $d_i=d_j$. However, in that case, the transformation $\Omega$ will leave $\mathcal{D}\eta$ invariant anyway: it will correspond to a trivial reshuffling of the equal $d_i$'s in $\mathcal{D}$. Indeed, if without loss of generality we arrange the elements of $\mathcal{D}$ in $k$ groups
\begin{equation}
\mathcal{D}=(d_1,\ldots,d_1,d_2,\ldots,d_2,\ldots,d_k,\ldots,d_k)~,
\end{equation}
$\omega$ will be block diagonal:
\begin{equation}
\omega=\begin{pmatrix}
\omega_1 & 0 & \cdots & 0 \\ 
0 & \omega_2 &  & \vdots \\ 
\vdots & • & \ddots & 0\\ 
0 & \cdots & 0& \omega_k
\end{pmatrix}~,
\end{equation}
and we get
\begin{equation}
\Omega(z)=\Omega_c\,\Omega_\omega(z)~,
\end{equation}
with $\Omega_c$ being constant and
\begin{equation}
\Omega_\omega(z)=\begin{pmatrix}
\Omega_1 & 0 & \cdots & 0 \\ 
0 & \Omega_2 &  & \vdots \\ 
\vdots & • & \ddots & 0\\ 
0 & \cdots & 0& \Omega_k
\end{pmatrix}~.
\end{equation}
This $z$-dependent part of the transformation acts however trivially on $\mathcal{D}$:
\begin{align}
\Omega \mathcal{D}\eta\Omega^T&=\Omega \mathcal{D}\Omega^{-1}\eta \\
&=\Omega_c\,\Omega_\omega(z)\mathcal{D}\Omega_\omega^{-1}(z)\Omega_c^{-1}\eta \\
&=\Omega_c\mathcal{D}\Omega_c^{-1}\eta \\
&=\Omega_c \mathcal{D}\eta\Omega^T_c~.
\end{align}
We see therefore that even when there are repeated eigenvalues, $\mathcal{M}$ can be diagonalised with a constant SO$(p,q)$ transformation throughout the flow.

\section{Jordan Normal Forms and \texorpdfstring{$\SO(p,q)$}{}} \label{App:Jordan}

Let us construct the basis for the bilinear form $\eta$ that makes the standard Jordan normal form $\mathcal{J}$ generalised symmetric. This standard form is given by
\begin{equation}
\mathcal{J}=\begin{pmatrix}
\mathcal{J}_1 & 0 & \cdots & 0 \\ 
0 & \mathcal{J}_2 &  & \vdots \\ 
\vdots & • & \ddots & 0\\ 
0 & \cdots & 0& \mathcal{J}_s
\end{pmatrix} \quad\text{with}\quad \mathcal{J}_i=\begin{pmatrix}
\lambda_i & 1 & 0 & \cdots & 0 \\ 
0 & \lambda_i & 1 & \ddots & \vdots \\ 
\vdots & • & \ddots & \ddots & 0 \\ 
• & • & • & \lambda_i & 1 \\ 
0 & • & \cdots & 0 & \lambda_i
\end{pmatrix} ~\,,
\end{equation}
then from the generalised symmetric condition we see that $\eta$ must be of the form
\begin{equation}
\eta=\begin{pmatrix}
\eta_1 & 0 & \cdots & 0 \\ 
0 & \eta_2 &  & \vdots \\ 
\vdots & • & \ddots & 0\\ 
0 & \cdots & 0& \eta_s
\end{pmatrix}~,
\end{equation}
where $\eta_i$ has the same size as the Jordan block to which it corresponds and is equal to $\pm 1$, if its size is 1, or 
\begin{equation}
\eta_i=\begin{pmatrix}
0 & \cdots & 0 & 1 \\ 
\vdots & • & 1 & 0 \\ 
0 & \iddots & • & \vdots \\ 
1 & 0 & \cdots & 0
\end{pmatrix} \,,
\end{equation}
otherwise.  Using all the previous, we can write \cite{Bergshoeff:2008be}
\begin{equation}
\mathcal{M}=\Omega \mathcal{J}\eta\Omega^{T}~\,,\qquad \Omega \in \SO(p,q)\,.
\end{equation}

We will now argue that $\mathcal{J}$ is strictly diagonal, and not just block diagonal. Since $\mathcal{M}=\mathcal{L}\mathcal{L}^T$ is positive definite, we must have that
$\mathcal{J} \eta=\Omega^{-1}\mathcal{M}(\Omega^{-1})^T$, is also positive definite.  Clearly, the matrix $\mathcal{J}\eta$ is block diagonal, and in order to be positive definite, each of its blocks $\mathcal{J}_i\eta_i$ must be positive definite. Suppose that there is at least one non-diagonal block $\mathcal{J}_i$ in the Jordan normal form for $\mathcal{A}$. The product $\mathcal{J}_i\eta_i$ is then
\begin{equation}
\mathcal{J}_i\eta_i=\begin{pmatrix}
0 & \cdots & 0 & 1 & \lambda_i \\ 
\vdots & • & 1 & \lambda_i & 0 \\ 
0 & \iddots & \iddots & \iddots & \vdots \\ 
1 & \lambda_i & 0 & • & • \\ 
\lambda_i & 0 & \cdots & • & 0
\end{pmatrix} ~.
\end{equation}
Defining the vector\footnote{We recall that $\lambda_i\neq0$ since det$(\mathcal{J})=$det$[\Omega^{-1}\mathcal{M}(\Omega^{-1})^T\eta]\neq 0$.} $v=(1,0,\ldots,0,-1/\lambda_i)$, we have
\begin{equation}
v^T\mathcal{J}_i\eta_i v=\left\{\begin{array}{cl}
-1 & \quad \text{if }\mathcal{J}_i \text{ has size }2\times 2~, \\ 
-2 & \quad \text{if }\mathcal{J}_i \text{ has size }r\times r \text{ with }r>2~,
\end{array}\right.
\end{equation}
from which we deduce that $\mathcal{J}_i\eta_i$ is not positive definite. We have hence shown by contradiction that $\mathcal{J}$ must be diagonal. To unify this discussion with the one of the previous subsection, we will write $\mathcal{D}=\mathcal{J}$. We know then that we can always diagonalize $\mathcal{M}$ using an SO($p,q$) transformation:
\be\label{normal1}
\mathcal{M}(z)= \Omega \mathcal{D}(z)\eta\Omega^T\,,
\ee
with $\mathcal{D}(z)$ diagonal and $\Omega\in \text{SO}(p,q)$. In this basis, $\eta$ takes the standard diagonal form.

\providecommand{\href}[2]{#2}\begingroup\raggedright\endgroup
\end{document}